\documentclass[a4paper,11pt,oneside]{article}

\usepackage[english]{babel}
\usepackage[T1]{fontenc}
\usepackage[utf8]{inputenc}

\usepackage[pdftex,unicode]{hyperref}
\hypersetup{pdftitle=Clustering Retail Products Based on Customer Behaviour}
\hypersetup{pdfauthor=Vladimír Holý and Ondřej Sokol and Michal Černý}

\usepackage{xcolor}
\definecolor{mycolor}{rgb}{0.16,0.00,0.50}
\hypersetup{colorlinks=true, linkcolor=mycolor, anchorcolor=mycolor, citecolor=mycolor, filecolor=mycolor, urlcolor=mycolor}

\usepackage[margin=60pt]{geometry}
\usepackage{amsmath}
\usepackage{amssymb}
\usepackage{bm}

\usepackage{enumitem}

\usepackage{graphicx}
\usepackage{float}
\usepackage{hhline}
\usepackage{multirow}

\usepackage[authoryear]{natbib}

\begin{document}

\begin{center}
{\Large \bfseries Clustering Retail Products Based on Customer Behaviour}
\end{center}

\begin{center}
{\bfseries Vladimír Holý} \\
University of Economics, Prague \\
Winston Churchill Square 4, 130 67 Prague 3, Czech Republic \\
\href{mailto:vladimir.holy@vse.cz}{vladimir.holy@vse.cz} \\
Corresponding Author
\end{center}

\begin{center}
{\bfseries Ondřej Sokol} \\
University of Economics, Prague \\
Winston Churchill Square 4, 130 67 Prague 3, Czech Republic \\
\href{mailto:ondrej.sokol@vse.cz}{ondrej.sokol@vse.cz}
\end{center}

\begin{center}
{\bfseries Michal Černý} \\
University of Economics, Prague \\
Winston Churchill Square 4, 130 67 Prague 3, Czech Republic \\
\href{mailto:cernym@vse.cz}{cernym@vse.cz}
\end{center}

\begin{center}
{\itshape October 1, 2016}
\end{center}

\noindent
\textbf{Abstract:}
The categorization of retail products is essential for the business decision-making process. It is a common practice to classify products based on their quantitative and qualitative characteristics. In this paper we use a purely data-driven approach. Our clustering of products is based exclusively on the customer behaviour. We propose a method for clustering retail products using market basket data. Our model is formulated as an optimization problem which is solved by a genetic algorithm. It is demonstrated on simulated data how our method behaves in different settings. The application using real data from a Czech drugstore company shows that our method leads to similar results in comparison with the classification by experts. The number of clusters is a parameter of our algorithm. We demonstrate that if more clusters are allowed than the original number of categories is, the method yields additional information about the structure of the product categorization.
\\

\noindent
\textbf{Keywords:} Product Categorization, Cluster Analysis, Genetic Algorithm, Retail Business, Drugstore Market.
\\

\noindent
\textbf{JEL Codes:} C38, M31.
\\

\section{Introduction}
\label{secIntro}

Categorization is important in retail business decision-making process. Product classification and customer segmentation belong to the most frequently used methods.
The customer segmentation is  focused on getting knowledge about the structure of customers and is used for targeted marketing.
For example \cite{Jonker2004} dealt with customer segmentation and its usability in marketing. Another approach to determining customer segmentation was used by \cite{Lockshin1997}. \cite{Seret2014} proposed a customer segmentation based on self-organizing maps with a priori prioritization in direct marketing.

The product categorization finds even more applications in marketing, e.g. new product development, optimizing placement of retail products on shelves, analysis of cannibalization between products and more general analysis of the affinity between products. \cite{Gruca2003} proposed a genetic algorithm to identify optimal new product position. A placement of retail products on shelves was studied by \cite{Borin1994}. Finding the right categories is also crucial for sales promotions planning. Cross-category sales promotion effect was studied in detail by \cite{Leeflang2008} and  \cite{Hruschka1999}.

Retail chains try to minimize costs everywhere. Among others their aim is to minimize the costs of product storage in stores. The storage management reaches the stage when stores often have no reserves in the drugstore storeroom because they are supplied dynamically two or more times per week. Therefore, it may happen that a store runs out of some products. The task is: 
\begin{enumerate}
\item How to fill a free place on shelves until the storage is restored.
\item How to find a product that best substitutes for the original one.
\end{enumerate}
Sold-out products are usually replaced by other ones from the same category, but it is not clear how to best define the categories from this viewpoint. 
This is the main business motivation behind this paper.

Products are almost always categorized according to their purpose, package properties, e.g. package size, brand and price level. However, there are different approaches to product categorization. For example \cite{Srivastava1981} used hierarchical clustering while \cite{Zhang2007} promoted fuzzy clustering. Another interesting possibilistic approach to clustering both customers and products was published by \cite{Ammar2016}.

Retail chains have available huge amount of market basket data, containing sets of items that a buyer acquires in one purchase, which can be used to efficiently model customer behaviour, e.g. \cite{Tsai2004}. However, these data are rarely taken into account in the product categorization. Data from market baskets are usually used for analysis of cross-category dependence for a priori given categories, e.g. \cite{Russell2000},  \cite{Mild2003}, \cite{Decker2003} and  \cite{Manchanda1999}. 

This paper proposes a new method for choosing categories utilizing market basket data.
Our method classifies products into clusters according to their common occurrences in the shopping baskets. 
Sets of products in individual shopping baskets as they were registered by the receipts are the only data used by  the method which assigns each product to just one category. The method determines product categories under given assumptions of product dependency in the same category.
It stems from the assumption that a customer buys only one product per category.
Experience shows that customers who buy one product from a given category  are generally less likely to buy also another product from the same category. The method applies a genetic algorithm to market basket data in order to find the best clusters of products based on their joint occurence in shopping baskets.

Retail companies usually inspect affinity relationship between single products, e. g. sales in the same basket normalised by total sales. However, clustering of products based solely on market basket data in this area is not so common. It can help mainly in organising shelf and/or maximising effect of promotional activities such as newsletter promotions with a significant discount. This kind of promotion should attract customers who do not regularly visit the store.
For example, the promotion of two products from the same category is not effective as customer usually buys only one of them. 
The interesting article focused on marketing strategies of associated products was published in \cite{Weng2016} in which author deals with the problem of assocation rules when product is marketed later. 
Our method for clustering may be helpful mainly in markets with the high proportion of sales in a promotion, such as Czech drugstore market where over half of sales is in the promotion. 

%ZAKLADNI METODY
As base method for clustering is usually taken $k$-means. Although $k$-means was proposed over 50 years ago, it is still one of the most used method. The overview of $k$-means and its modification can be found in \cite{Jain2010}. 

The overview of methods for automatic clustering using nature-based metaheuristic methods such as genetic algorithms or swarm intelligence can be found in \cite{Jose-Garcia2016}. Interesting approach using fuzzy chromosone in a genetic algorithm was published by \cite{Yang2015}. Another possibilistic approach was presented by \cite{Ammar2015}. Combination of $k$-means and ant colony optimization was published in \cite{Niknam2010}. 
 
The objective of this paper is to present a new method of retail product clustering based on shopping behaviour of customers. 
%highlight differences
Usually products which are commonly bought together are clustered into the same cluster. In our method we use different point of view. Our goal is to find clusters that minimize the number of occurences in which two products from the same cluster are in the same shopping basket. This is the main and the most important difference of our method.

The resulting categorization can be used not only for choosing the products suitable to replace sold-out ones but also for optimizing placement of retail products on shelves or for maximizing profit of sales promotions. 
To maximize the profit it is more effective to spread promotion across different categories instead of stacking multiple promotions in the same category.
The resulting clustering can also help in persuading customer into buying more expensive alternative, e.g. promotion which includes discount on more expensive product when a product from same category is bought.

The article is organized as follows: In Section \ref{secMeth}, we present the general idea of the proposed method and its assumptions. In Section \ref{secSim}, we test the method using synthetic data to illustrate its performance. We also show how the violation of the assumptions affects the method's results. In Section \ref{secData} the application to drugstore's market basket is presented and its potential to detect clusters in real data sets which have not been found before is demonstrated. The paper concludes with a summary in Section \ref{secCon}.

\section{Methods}
\label{secMeth}

In this section we propose a new method for clustering retail products based on customer behaviour. We also present our approach to evaluate resulting clustering.

To clarify terminology in this article we use \textit{categories} meaning the original product category that was defined expertly based on the character and the purpose of the products. On the other hand, \textit{clusters} are results of our method. Clusters are determined using only market basket data.

\subsection{Clustering Using Genetic Algorithm}
\label{secMethClu}

We formulate clustering of retail products as an optimization problem. The goal is to find clustering that minimizes the number of products within the same cluster in one shopping basket. It is based on the idea that in general customers will not buy more than one product from each cluster (products in clusters are similar so they need only one). We define a cost function which penalizes a violation of this behaviour. For given clustering the cost function calculates weighted number of violations of the assumption that in each basket there is at most one product from a cluster.

We approach clustering as a series of decisions. For each pair of products there is a decision whether these two products should be in the same cluster or not. It is inspired by the Rand index (formulated later in Subsection \ref{secMethEva}). Specifically, we minimize the average ratio of incorrect clustering decisions. Here, incorrect is meant in the sense that they lead to multiple products within the same cluster in one shopping basket.

Let $n_B$ be the number of baskets, $n_P$ the number of products and $n_C$ the maximum number of clusters. We define matrix $\bm{A}$ with $n_B$ rows, $n_P$ columns and elements $a_{i,j}$ as
\begin{equation}
a_{i,j} = \begin{cases}
1 & \text{if product } j \text{ is present in basket } i, \\
0 & \text{otherwise}.
\end{cases}
\end{equation}
A possible clustering is defined as $\bm{x}=(x_1,\ldots,x_{n_P})'$, where $x_j$ is an integer and $1 \leq x_j \leq n_C$. Elements of vector $\bm{x}$ correspond to products and their values represent assignment of a product to a cluster.

For each basket $b=1,\ldots,n_B$ we calculate the total number of decisions $D_b$ as
\begin{equation}
\label{eqCost2}
D_b=\binom{d_b}{2}, \qquad d_b=\sum_{j=1}^{n_P}a_{b,j}
\end{equation}
and the number of decisions that lead to multiple products within the same cluster $V_b$ as
\begin{equation}
\label{eqCost1}
V_b (\bm{x})= \sum_{c : v_{b,c}(\bm{x})>1} \binom{v_{b,c}(\bm{x})}{2}, \qquad v_{b,c}(\bm{x})=\sum_{j:x_j=c}a_{b,j}.
\end{equation}
The number of violating decisions $V_b$ is dependent on clustering $\bm{x}$. Finally, we define the cost function as
\begin{equation}
\label{eqCost}
f_{cost} (\bm{x}) = \frac{1}{n_B} \sum_{b=1}^{n_B} \frac{V_b(\bm{x})}{D_b}.
\end{equation}
Hence, the cost function equals to the average ratio of decisions in which two products from the same cluster are in the same shopping basket. The range of the cost function is from 0 to 1. If there is no basket containing products from the same cluster, then the cost function is 0. On the other hand, if every basket contain only products from the same cluster, then the cost function is 1.   

We have considered several other objective function formulations, most notably:
\begin{itemize}
\item the ratio of total number of multiple products within the same cluster,
\item the average ratio of multiple products within the same cluster over all baskets,
\item the average ratio of multiple products within the same cluster over all products,
\item the ratio of total number of baskets with multiple products within the same cluster,
\end{itemize}
but the best clusterings were given by the objective function \eqref{eqCost} inspired by Rand index.

The whole optimization problem is formulated as
\begin{equation}
\label{eqOpti}
\begin{aligned}
\min_{\bm{x}} \quad & f_{cost} (\bm{x}) \\
\text{s. t.} \quad & x_i \leq n_C & \text{for } i=1,\ldots,n_P, \\
& x_i \in \mathbb{N} & \text{for } i=1,\ldots,n_P. \\
\end{aligned}
\end{equation}
The cost function \eqref{eqCost} is minimized over all possible clusterings $\bm{x}$. The maximum number of clusters $n_C$ is a fixed number. If we did not limit the number of clusters, each product would be assigned to its own cluster. Optimization problem \eqref{eqOpti} is an integer non-linear programming, which can easily be shown to be NP-hard. To efficiently solve this or at least to get approximate solution we use heuristic \textit{genetic algorithm}. Details of our genetic algorithm parameters are discussed in Subsection \ref{secSimParam}. Genetic algorithm for solving an integer non-linear program was already used by \cite{Jiang1997} and more recently by \cite{Yang2015}.

To ensure that resulting clustering is meaningful we need to make following assumptions:
\begin{enumerate}
\item[(A1)] The probability that customer buys at least two products from one category is strictly less than 50 \%.
\item[(A2)] The true number of clusters is known.
\item[(A3)] Each customer has the same nonzero probability of buying a specific product and the probability is constant in time.
\end{enumerate}
Assumption (A1) tells us that although our model allows customers to buy more than one product within the same cluster in one shopping basket, this behaviour is not considered standard but as a model error. Assumption (A2) reflects formulation of our optimization problem in which we specify the maximum number of clusters. In almost all cases the resulting number of clusters is the maximum number of clusters (more clusters are preferred by the objective function). Finally, Assumption (A3) is meant to prevent situations in which some customers buy product \textit{A} and never product \textit{B} while other customers buy product \textit{B} and never product \textit{A}. This behaviour would result in assigning products \textit{A} and \textit{B} into the same cluster even if they are completely different. Assumption (A3) ensures that with a large enough dataset, all combinations of products will appear in some baskets with probability approaching 1. Later, in Section \ref{secSim} we discuss in more detail how violation of these assumptions would affect the resulting clustering.

\subsection{Evaluation of Clustering}
\label{secMethEva}

We use three different statistics to evaluate our resulting clustering when true categories are known. The first one is \textit{purity} used for example by \cite{Manning2008}. It is computed in a very straightforward way. Each estimated cluster is assigned a category which is the most frequent in the cluster. Purity is then the ratio of products with correctly assigned categories. It is calculated as
\begin{equation}
I_{PUR} = \frac{1}{n_P} \sum_{i=1}^{n_C} \max_{j} | C_i \cap R_j |,
\end{equation}
where $n_P$ is the number of products, $n_C$ is the number of estimated clusters, $C_i$ is the set of products in estimated cluster $i$ and $R_j$ is the set of products in real category $j$. Bad clusterings have the purity close to 0, the perfect clustering has purity equal to 1. However, if the number of estimated clusters is much larger than the number of real categories the purity always has a high value and in this case it is not very meaningful.

For this reason we also use a modification of the purity in which we reverse the role of true categories and estimated clusters. \textit{Reverse purity} is defined as
\begin{equation}
I_{REV} = \frac{1}{n_P} \sum_{j=1}^{n_R} \max_{i} | C_i \cap R_j |,
\end{equation}
where $n_R$ is the number of real categories.

%With more approximately equally sized clusters than original approximately equally sized categories the statistics becomes limited from above. For example if we have 100 products in 10 equally sized categories and 20 equally sized clusters, than the purity is limited from above by 0.5 as each category contain maximum of 5 products from the same cluster. Thefore, purity as well as reverse purity in case of different number of categories and clusters is not appropriate evaluation statistic. 

The last statistic we use is the \textit{Rand index} proposed by \cite{Rand1971}. It is based on the idea that clustering is a series of decisions that either put two products into the same cluster or put them to different clusters. Therefore the number of decisions is the number of product pairs. Rand index is defined as a ratio of correct decisions and is calculated as
\begin{equation}
I_{RAND} = \frac{P_{TP}+P_{TN}}{P},
\end{equation}
where $P_{TP}$ is the number of pairs correctly assigned to the same cluster, $P_{TN}$ is the number of pairs correctly assigned to different clusters and $P$ is the number of all pairs. Accurate clusterings have Rand index close to 1.

\section{Simulation Study}
\label{secSim}

To reveal properties of our method we perform several simulations. In Subsection \ref{secSimParam} we compare different parameters of our genetic algorithm. In three remaining subsections we simulate a violation of assumptions (A1), (A2) and (A3). In all simulations we consider a dataset consisting of 10 000 shopping baskets, each with 4 categories that have one or two products. We have a total of 10 categories with 10 products each. We simulate the behaviour of a customer who selects a set of categories (s)he needs (sets of 4 categories are selected with the same probability, apart from Subsection \ref{secSimCustomer}). Then he selects which products from that category he wants to buy (products are selected with the same probability and there is also a 10\% chance to buy two products instead of one). We have chosen this characteristics of simulated data because they are similar to the size and the structure of the real dataset we use later in Section \ref{secData}.

\subsection{Choosing Genetic Algorithm Parameters}
\label{secSimParam}

To properly use genetic algorithm we need to choose several parameters. The first one is the \textit{size of population}. With bigger population of individuals more possible clusterings are explored, which can result in finding better solution. We set the population size to 500 individuals due to computational complexity.

Initial population is generated randomly and then the algorithm iteratively selects new populations. The number of iterations is another parameter called the \textit{number of generations}. We always use the fixed number of 1 000 generations. However, our simulations show that most clusterings converge much faster. Each candidate solution is represented by an individual. Properties of candidate solutions are encoded in chromosomes of individuals. At each generation a percentage of individuals with lowest values of cost function (called the elite population) passes to the next generation without any alteration. We consider the \textit{ratio of elite population} from 0 to 0.2. In our case elite population does not have big impact on the best individual in last generation, all values result in perfect or almost perfect clustering. It is meaningful to carry over at least the best individual from previous generation so the quality of solution will not decrease. We set the elite ratio to 0.1.

\begin{figure}
\begin{center}
\includegraphics[width=11cm]{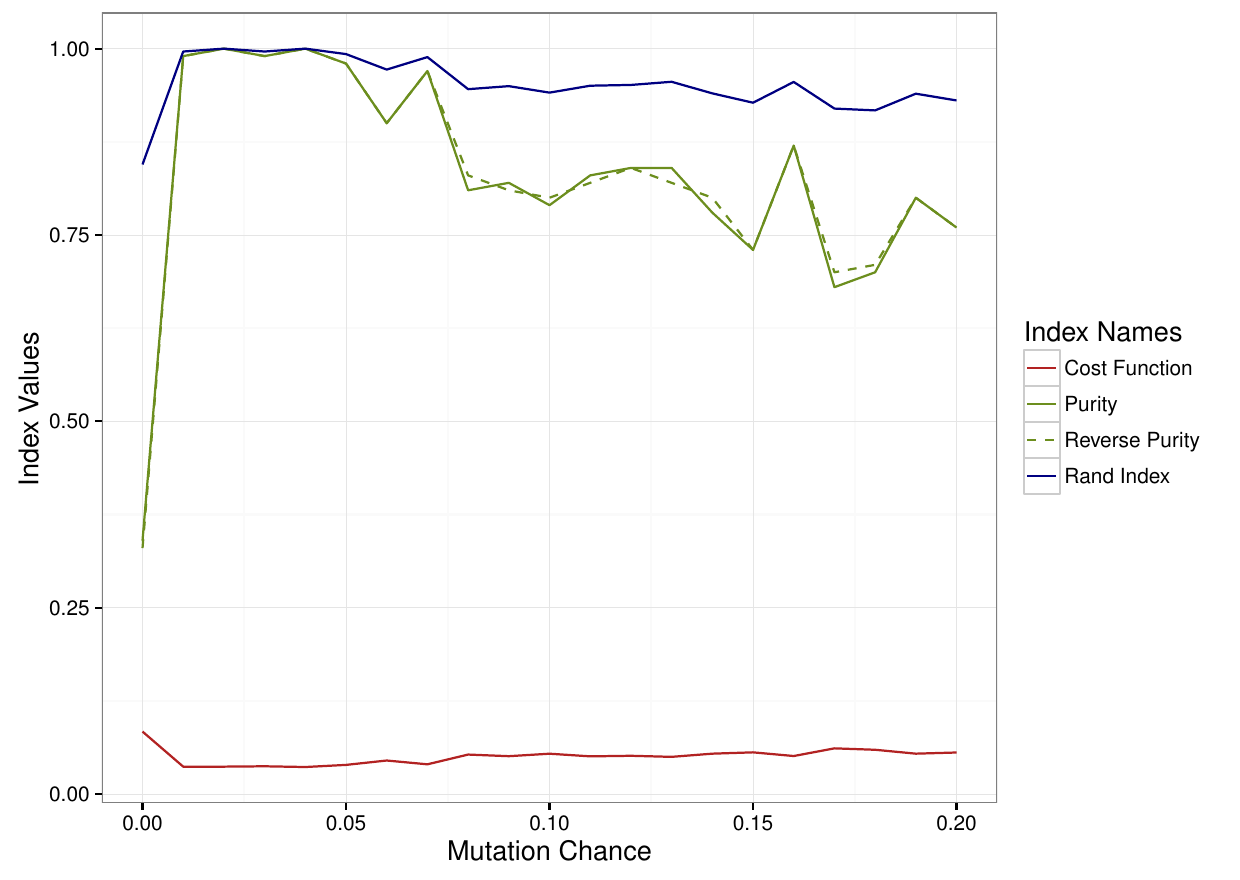} 
\caption{Evaluation statistics for genetic algorithm with different mutation chances.}
\label{figMutations}
\end{center}
\end{figure}

Chromosomes of the rest of the new individuals are generated by crossover and mutation operations. For each new individual (called the child) crossover selects two individuals from the last genration (called the parents). The parents are selected randomly with weights according to the sorting by their cost function. The child is then created by combining chromosomes of both parents. We use one-point crossover which means that a random number of chromosomes $c$ is selected. Then the first $c$ chromosomes of the child are taken from one parent while the remaining chromosomes are taken from the other parent. Finally the mutation operation is performed. Each chromosome of the child has a probability of changing its value to a random one. This probability is the last parameter called the \textit{mutation chance}. We consider this parameter to be from 0 to 0.2. Figure \ref{figMutations} shows that positive mutation chances up to 0.04 result in perfect or almost perfect clustering. Simulation with no mutation chance gives far worse result showing the importance of mutation. We set the mutation chance to 0.01 to allow algorithm to concentrate on improving one point while retaining some exploratory ability of mutation.

\begin{figure}
\begin{center}
\includegraphics[width=11cm]{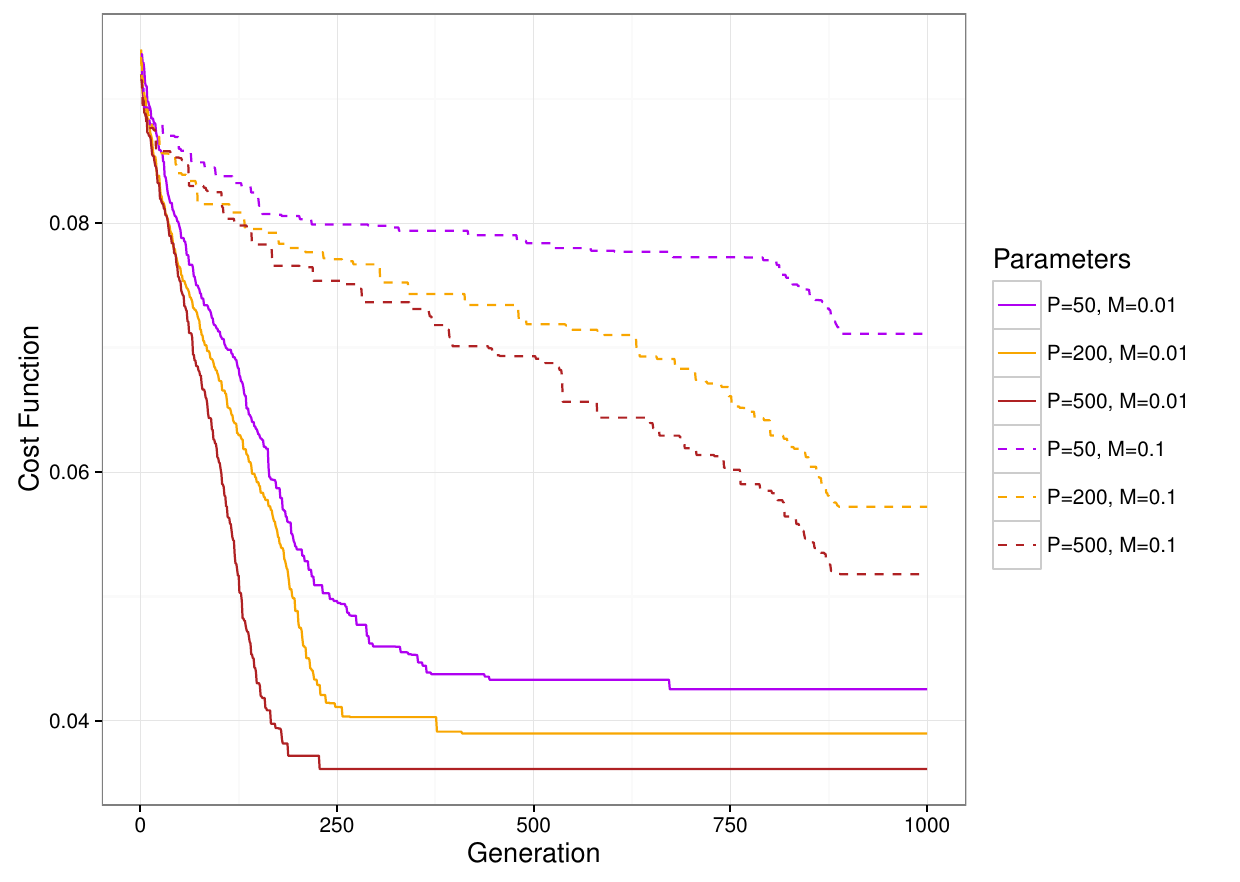} 
\caption{Cost function of best individuals in each generation for different population sizes $P$ and mutation chances $M$.}
\label{figGeneration}
\end{center}
\end{figure}

Next, we analyze combinations of the mentioned parameters. In Figure \ref{figGeneration} cost functions of the best individuals in each generation are shown for 6 different parameter settings. We combine population sizes $P=(50,200,500)$ with mutation chances $M=(0.01,0.1)$ while the ratio of elite population is set to $0.1$. In Table \ref{tabParam} several statistics of the resulting clustering are presented. We can see that lower mutation chance leads to faster convergence. There is a risk of lower mutation chance to end up in a local minimum but the results show this is not the case. The population size of 500 individuals with mutation chance of 1\% resulted in a perfect clustering. In the rest of the paper, these are the genetic algorithm parameters we use.

\begin{table}
\begin{center}
\begin{tabular}{lcccccc}
\hline
Population Size & 50 & 200 & 500 & 50 & 200 & 500 \\
Mutation Chance & 0.01 & 0.01 & 0.01 & 0.1 & 0.1 & 0.1 \\ \hline
Cost Function & 0.043 & 0.039 & 0.036 & 0.071 & 0.057 & 0.052 \\
Purity & 0.940 & 0.980 & 1.000 & 0.550 & 0.740 & 0.081 \\
Reverse purity & 0.940 & 0.980 & 1.000 & 0.550 & 0.750 & 0.081 \\
Rand index & 0.978 & 0.993 & 1.000 & 0.886 & 0.930 & 0.946 \\ \hline
\end{tabular}
\caption{Statistics of the best individual in the final generation for different population sizes $P$ and mutation chances $M$.}
\label{tabParam}
\end{center}
\end{table}

\subsection{Multiple Products Within the Same Category in One Shopping Basket}
\label{secSimSecond}

\begin{figure}
\begin{center}
\includegraphics[width=11cm]{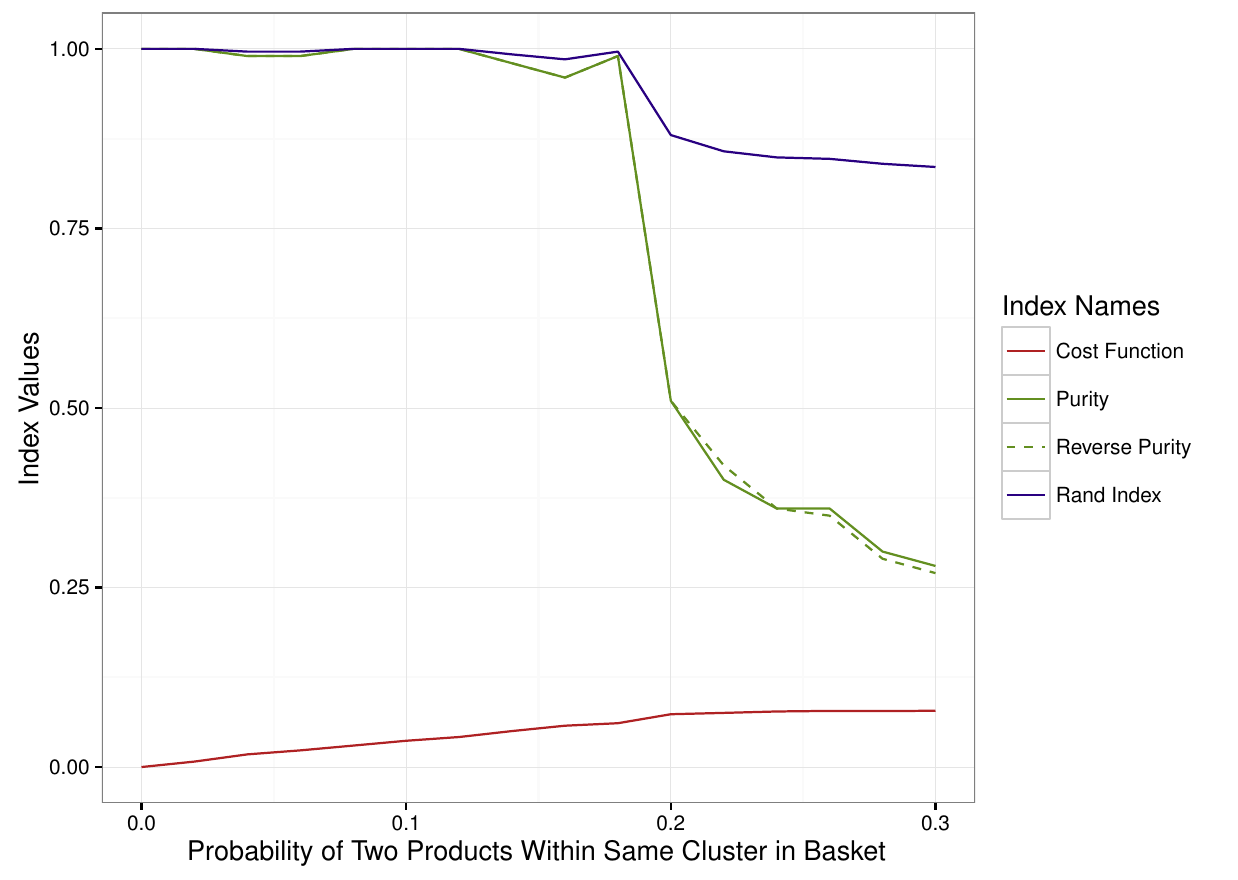} 
\caption{Evaluation statistics for clustering data with different probabilities of second product in the same category in one shopping basket.}
\label{figSecond}
\end{center}
\end{figure}

In this subsection we study the sensitivity of the proposed method to the situation, in which customers buy more than one product from the same category. We simulate data for different probabilities of buying the second product. Results are shown in Figure \ref{figSecond}. As we can see the method gives almost perfect clustering for the probability of the second product up to 0.18. At probability 0.20 there is a significant decrease in accuracy. This is caused by the loss of relevant information contained in shopping basket data supplied to the cost function. If we could increase the number of observed shopping baskets or the average number of products in a shopping basket we would get more precise results even for the second product probability of 0.20 or higher.

\subsection{Unknown Number of Clusters}
\label{secSimNumber}

\begin{figure}
\begin{center}
\includegraphics[width=11cm]{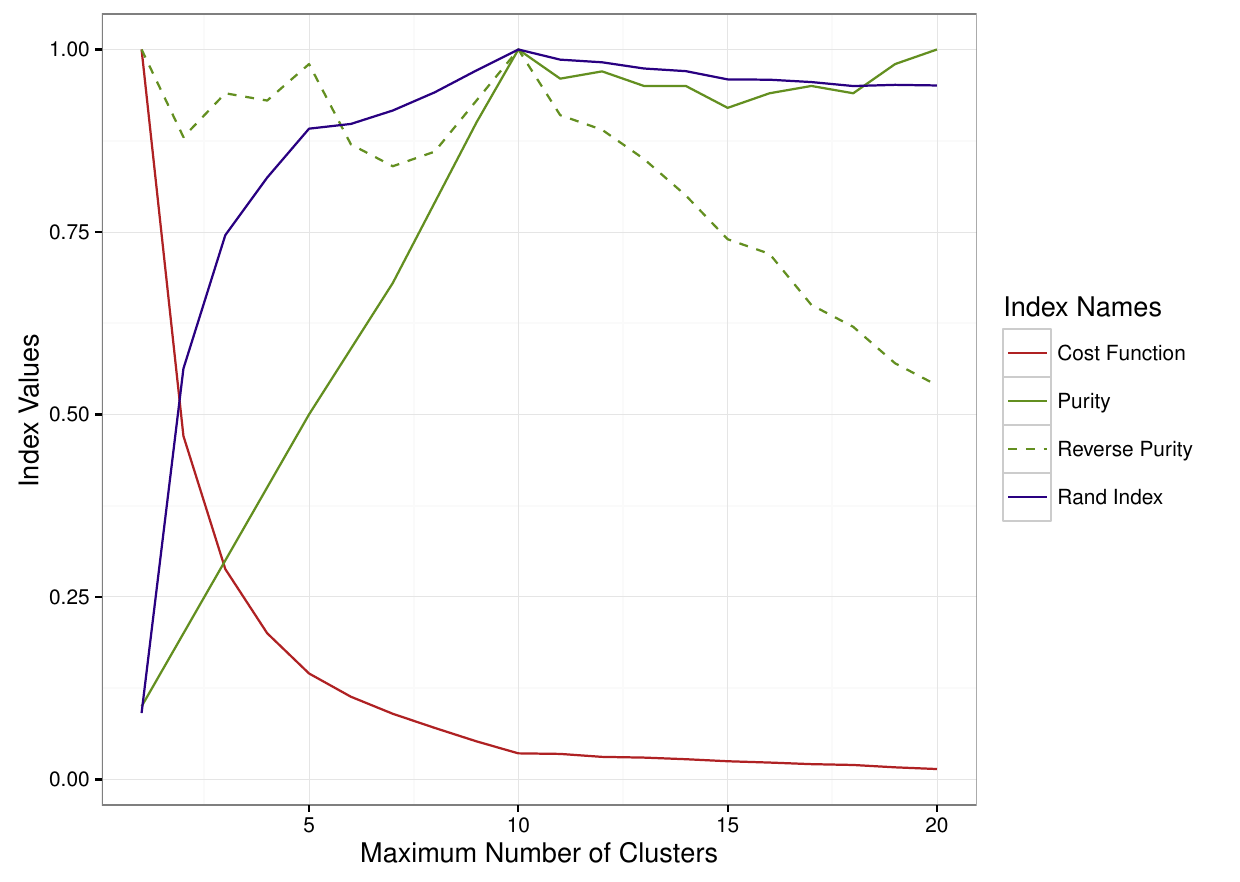} 
\caption{Evaluation statistics for clustering when number of categories is unknown.}
\label{figNumber}
\end{center}
\end{figure}

We have assumed in our simulations so far that the true number of categories is known. Now we inspect the behaviour of our method when used with different numbers of clusters. Results are shown in Figure \ref{figNumber}. The question is if we can identify the correct number of categories. In Figure \ref{figNumber} we can see that the purity, the reverse purity and the Rand index have value of 1 for 10 clusters, indicating the perfect clustering. However, in a real application we do not know the true categorization and therefore we cannot calculate the purity statistics or the Rand index. A way to determine what number of clusters should be used is to analyse the shape of the cost function. For a number of clusters less than the true number the cost function is significantly decreasing with more clusters. When the true number of categories has been reached the cost function continues to decrease only by a small amount. As we can see in Figure \ref{figNumber} the cost function stops rapidly decreasing around 10 clusters which is the true number of categories.

\subsection{Different Types of Customers}
\label{secSimCustomer}

\begin{figure}
\begin{center}
\includegraphics[width=11cm]{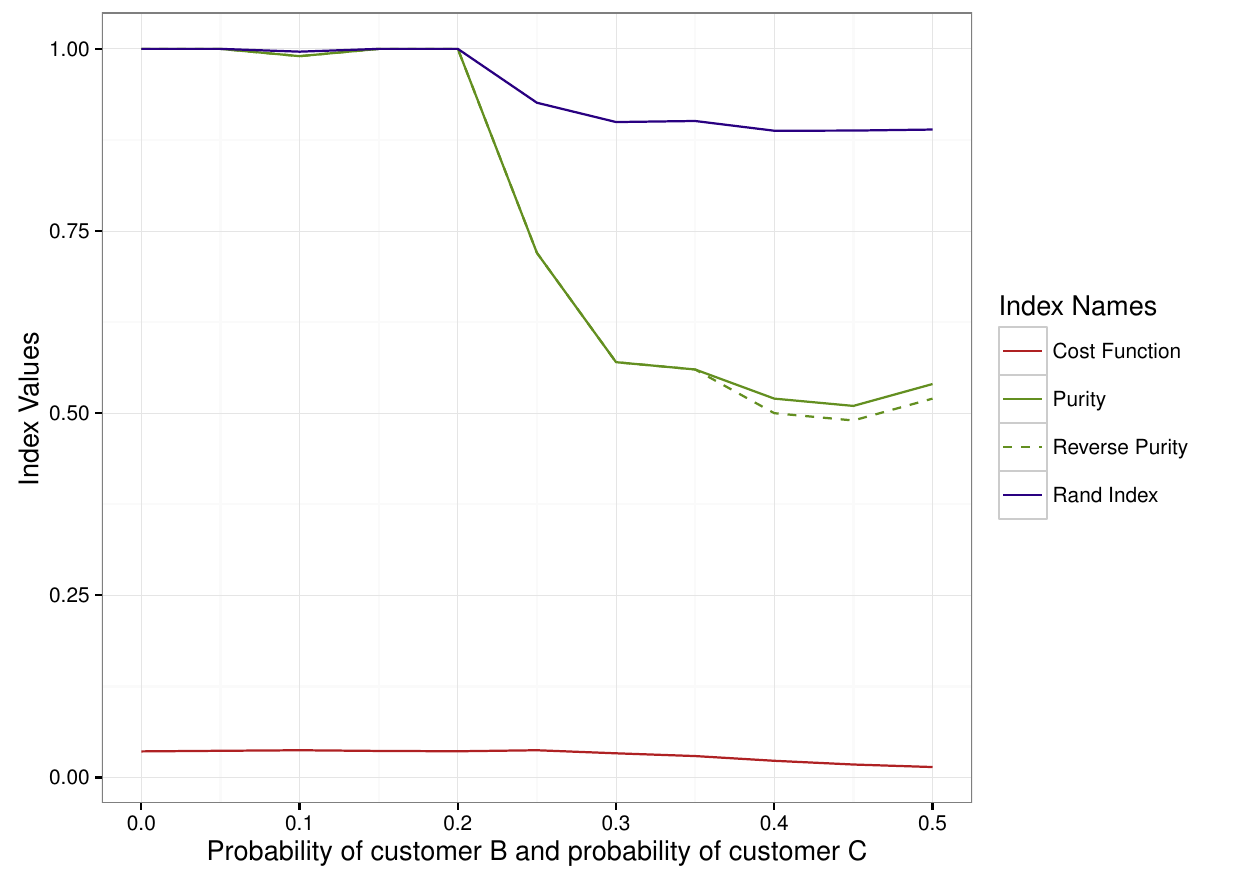} 
\caption{Evaluation statistics for clustering data with different probabilities of occurrence of customers \textit{B} and \textit{C}.}
\label{figCustomer}
\end{center}
\end{figure}

Finally, we discuss a violation of Assumption (A3). We consider three types of customers. Customer \textit{A} can buy products from all categories with equal probability. Customer \textit{B} can buy products only from a half of categories while customer \textit{C} can buy products only from the other half of categories. We study the behaviour of the proposed method for customer structures ranging from all customers being of type \textit{A} (this was the case of all previous simulations) to half customers being type \textit{B} and half type \textit{C}. Results are shown in Figure \ref{figCustomer}. If the customers violating assumption (A3) are in minority the resulting clustering is not affected. However, from the point where customer composition is 50\% type \textit{A}, 25\% type \textit{B} and 25\% type \textit{C} the resulting clustering becomes quite chaotic.

\section{Data Analysis}
\label{secData}

In this section, we use our method with a sample of real data. Our dataset consists of individual purchase data of one of the retail chains in drugstore market in the Czech Republic. We take original categories applied in the retail chain that are defined expertly, according to the character and the purpose of the products, as a reference classification.

Customer behaviour in the Czech drugstore market is specific. As there is a~high density of malls and hypermarkets, customers who visit drugstores usually buy just a few  products. The average number of items in a shopping basket in the drugstore is around 3.

\subsection{Description of Dataset}
In this paper we use a sample of receipts containig at least 4 products from the whole year 2015. 
The size of our dataset is 10 608 baskets containing 10 best-selling products from 10 most-popular categories defined by the drugstore. Original categories are based mostly on product purpose and price level.
Thus, we have 100 different products.
 The sample is mainly for illustration how our method work and how the clusters are defined -- that is the reason why we have chosen exactly 100 products.  

Parameters of the genetic algorithm are the same as presented in Section \ref{secSimParam}.

To clarify the terminology in this section, original categories are denoted by letters while clusters found by our method are denoted by numbers.

\subsection{Model with 10 Clusters}

We applied the proposed method with the maximum number of 10 clusters and we compared the found clusters with the original categories. Using the real dataset we have found that Assumption (A1) about consumers' behaviour was violated in approximately $19 \%$ of baskets on average. The ratios of violations significantly differ for each category as shown in Table \ref{tabViolation}. 
%As the second column in Table \ref{tabViolation} the assumption (A3) was also violated because product categories are not equally popular. 

The assignment of products to clusters is shown in Table \ref{tabDataAss10}. It is apparent that the proposed method had a problem with assigning products from categories \textit{C, D} and \textit{E}. Those are the categories with the highest percentage of violations of Assumption (A1). 

The method produced 10 clusters which was the maximum allowed. Evaluation statistics of the results of this test are shown in Table \ref{tabDataEval10}. Purity as well as reverse purity statistics show that some of the products were not assigned as in original categorization. The cost function value of the assignment is $0.0153$ which is lower than the value of the expert estimate assignment which is $0.0182$. The reason is that the method minimizes the number of products bought together within one cluster. Therefore, if a category suffers from a violation of Assumption (A1), our method puts together products from different original categories into one cluster to minimize the cost function.

Using our method we have found out that categories \textit{C, D} and \textit{E} are perceived differently by the customers and by the management. This finding can be further used in designing new product categorization or in defining subcategories.

\begin{table}
\begin{center}
\begin{tabular}{clcc}
\hline
Category & Name & Occurences  & Violation ratio \\ \hline
\textit{A} & Dishwashing liquid & 4387& 0.022        \\
\textit{B} & WC liquid cleaners       &4212& 0.047        \\
\textit{C} & Handkerchieves and napkins & 5769& 0.077 \\
\textit{D} & Soap & 2026& 0.179 \\
\textit{E} & Tampons      &1837& 0.104        \\
\textit{F} & Toilet paper      &6993& 0.020        \\
\textit{G} & Trash bags       &4543& 0.068        \\
\textit{H} & Paper towels      &4544& 0.012        \\
\textit{I} & Cotton wool and cotton buds      &3991& 0.026        \\
\textit{J} & Facial pads     &5224& 0.012         \\ \hline
\end{tabular}
\caption{Violations for each category.}
\label{tabViolation}
\end{center}
\end{table}

\begin{table}
\begin{center}
\begin{tabular}{cccccccccc}
\hline
\multicolumn{10}{c}{Original categories} \\
\textit{A} & \textit{B} & \textit{C} & \textit{D} & \textit{E} &
\textit{F} & \textit{G} & \textit{H} & \textit{I} & \textit{J} \\ \hline
7 & 6 & 8 & 3 & 5 & 1 & 4 & 3 & 9 & 10 \\
7 & 5 & 2 & 2 & 5 & 1 & 4 & 3 & 9 & 10 \\
7 & 6 & 8 & 2 & 5 & 1 & 4 & 3 & 9 & 10 \\
7 & 6 & 8 & 5 & 6 & 1 & 4 & 3 & 9 & 10 \\
7 & 6 & 2 & 5 & 5 & 1 & 4 & 3 & 9 & 10 \\
7 & 6 & 8 & 2 & 5 & 1 & 5 & 3 & 9 & 10 \\
7 & 6 & 8 & 2 & 7 & 1 & 4 & 3 & 9 & 10 \\
7 & 6 & 8 & 2 & 5 & 1 & 4 & 3 & 9 & 10 \\
7 & 6 & 8 & 5 & 5 & 1 & 4 & 5 & 9 & 10 \\
7 & 6 & 2 & 2 & 5 & 1 & 4 & 3 & 5 & 10 \\ \hline
\end{tabular}
\caption{Assignment of 100 products from original categories \textit{A}-\textit{J} to clusters 1--10.}
\label{tabDataAss10}
\end{center}
\end{table}

\begin{table}
\begin{center}
\begin{tabular}{lc}
\hline
Number of classes & 10 \\ 
Purity     & 0.870        \\
Reverse purity       & 0.870        \\
Rand index        & 0.955        \\
Cost function value       & 0.0153        \\ \hline
\end{tabular}
\caption{Evaluation statistics for model with 10 clusters.}
\label{tabDataEval10}
\end{center}
\end{table}

\subsection{Model with 8 Clusters}

In our next test we assign the same 100 products of 10 categories into 8 clusters. The results are shown in Table \ref{tabDataAss8}. There is a good correspondence between 8 clusters and 8 categories, \textit{A, B, C, F, G, H, I} and \textit{J}. Products from the \textit{problematic} categories \textit{D} and \textit{E} were assigned quite randomly to clusters 2 to 8. Evaluation statistics of this model are in Table \ref{tabDataEval8}. Results confirmed that categories \textit{A, B, C, F, G, H, I} and \textit{J} are perceived similary by the custorers and by the managers. 
As expected, purity statistic is lower than in the test of Section \ref{tabDataAss10} with more clusters and the cost function has higher value. Purity has to be lower as the size of categories is generally larger than the size of clusters if the cost function is minimized.

\begin{table}
\begin{center}
\begin{tabular}{ccccccccccc}
\hline
\multicolumn{10}{c}{Original categories} \\
\textit{A} & \textit{B} & \textit{C} & \textit{D} & \textit{E} & \textit{F} & \textit{G} & \textit{H} & \textit{I} & \textit{J} \\ \hline
8          & 5          & 2          & 4          & 3          & 1          & 7          & 3          & 6          & 4          \\
8          & 5          & 2          & 6          & 8          & 1          & 7          & 3          & 6          & 4          \\
8          & 5          & 2          & 3          & 5          & 1          & 7          & 3          & 6          & 4          \\
8          & 5          & 2          & 2          & 3          & 1          & 7          & 3          & 6          & 4          \\
8          & 5          & 2          & 5          & 8          & 1          & 7          & 3          & 6          & 4          \\
8          & 5          & 2          & 3          & 5          & 1          & 7          & 3          & 6          & 4          \\
8          & 5          & 2          & 6          & 8          & 1          & 7          & 3          & 6          & 4          \\
8          & 5          & 2          & 4          & 5          & 1          & 7          & 3          & 6          & 4          \\
8          & 5          & 2          & 2          & 7          & 1          & 7          & 4          & 5          & 4          \\
8          & 5          & 5          & 6          & 7          & 1          & 7          & 3          & 6          & 4          \\ \hline
\end{tabular}
\caption{Assignment of 100 products from original categories \textit{A}-\textit{J} to clusters 1-8.}
\label{tabDataAss8}
\end{center}
\end{table}

\begin{table}
\begin{center}
\begin{tabular}{lc}
\hline
Number of classes & 8 \\ 
Purity     & 0.770        \\
Reverse purity      & 0.830        \\
Rand index        & 0.931        \\
Cost function value       & 0.027        \\ \hline
\end{tabular}
\caption{Evaluation statistics for model with 8 clusters.}
\label{tabDataEval8}
\end{center}
\end{table}

\subsection{Model with 13 Clusters}
In this model we tried to assign 100 products from 10 categories into 13 clusters -- a little more clusters than the number of given categories. The resulting assignment is shown in Table \ref{tabDataAss13}. 

The method created cluster 4 which contains products of 3 different categories. Again we can see that category \textit{D} (soap), which has the highest violation ratio, tends to be split up. On the other hand, category  \textit{J} (facial pads) which has the lowest violation ratio remains the same. Category \textit{G} (thrash bags) is split up into two exclusive clusters. That makes sense as this category includes both thick and thin thrash bags. It is apparent that categories \textit{C}, \textit{D} and \textit{E} were splitted into more clusters. Therefore customers buying items from these categories are more likely to buy more different products within the same category. This finding could help in planning promotions where customer gets discount on more expensive product when a product from same category is bought -- those promotions are more effective for clusters which are not splitted.

Evaluation statistics are shown in Table \ref{tabDataEval13}. 
Reverse purity statistics is lower than in the previous cases. That is expected result as we estimated more categories than the number of the original ones.

\begin{table}
\begin{center}
\begin{tabular}{cccccccccc}
\hline
\multicolumn{10}{c}{Original categories} \\
\textit{A} & \textit{B} & \textit{C} & \textit{D} & \textit{E} &
\textit{F} & \textit{G} & \textit{H} & \textit{I} & \textit{J} \\ \hline
11&9&13&1&6&10&2&8&12&3\\
11&1&5&4&6&10&2&8&12&3\\
11&9&13&6&4&10&7&8&12&3\\
11&9&5&13&7&10&2&8&12&3\\
11&9&5&1&6&10&2&4&12&3\\
11&9&5&6&4&10&7&8&12&3\\
11&9&13&4&4&1&7&8&12&3\\
6&9&5&1&4&10&2&8&4&3\\
11&9&5&1&4&10&7&4&12&3\\
6&9&5&4&1&10&7&8&2&3\\ \hline
\end{tabular}
\caption{Assignment of 100 products from original categories \textit{A}-\textit{J} to clusters 1--13.}
\label{tabDataAss13}
\end{center}
\end{table}

\begin{table}
\begin{center}
\begin{tabular}{lc}
\hline
Number of classes & 13 \\ 
Purity     &  0.840        \\
Reverse purity      & 0.730        \\
Rand index        & 0.946        \\
Cost function value       & 0.0083        \\ \hline
\end{tabular}
\caption{Evaluation statistics for model with 13 clusters.}
\label{tabDataEval13}
\end{center}
\end{table}

\subsection{Model with 20 Clusters}

We have shown that the proposed method can determine categories which were originally defined expertly based on the nature of the products if the assumption (A1) is not significantly violated. In this test we assign products to 20 clusters. The resulting assignment is shown in Table \ref{tabDataAss20}.

From Table \ref{tabDataEval20} it follows that categories \textit{A, B, C, F, G, H, I} and \textit{J} were split into two or three clusters which can be used to define subcategories. Conversely, the categories \textit{D} and \textit{E} contain more clusters. None of these clusters are limited only to a single category.
Categories \textit{D} and \textit{E} violate the assumption (A1) more than other categories. On the other hand the method made some interesting and reasonable clusters. For example in category \textit{B} all four WC liquid cleaners were clustered with pine aroma. That leads us to a fact that customers usually do not buy more liquid cleaners with the same aroma. Hence, products with the same function and aroma should be placed next to each other instead of sorting by brand as customers are choosing one within the products with the same aroma. 

Some other clusters can not be easily described. Finding not so obvious clusters is the advantage of our method. 

Evaluation statistics are shown in Table \ref{tabDataEval20}. 
Reverse purity statistics is again significantly lower as we assign to more clusters. 
The cost function value is also significantly lower than in the model of \ref{tabDataAss10} as expected.

\begin{table}
\begin{center}
\begin{tabular}{cccccccccc}
\hline
\multicolumn{10}{c}{Original categories} \\
\textit{A} & \textit{B} & \textit{C} & \textit{D} & \textit{E} &
\textit{F} & \textit{G} & \textit{H} & \textit{I} & \textit{J} \\ \hline
19 & 4  & 11 & 18 & 18 & 3 & 5  & 16 & 17 & 13 \\
10 & 15 & 8  & 18 & 15 & 1 & 14 & 20 & 2  & 13 \\
19 & 7  & 6  & 17 & 12 & 3 & 9  & 20 & 2  & 13 \\
19 & 7  & 6  & 11 & 15 & 1 & 14 & 16 & 17 & 13 \\
10 & 4  & 8  & 12 & 18 & 3 & 5  & 20 & 2  & 13 \\
10 & 7  & 8  & 15 & 12 & 3 & 14 & 16 & 17 & 13 \\
10 & 4  & 11 & 12 & 5  & 1 & 9  & 16 & 2  & 13 \\
10 & 4  & 8  & 12 & 18 & 1 & 14 & 20 & 2  & 4  \\
10 & 7  & 8  & 18 & 5  & 3 & 9  & 15 & 11 & 19 \\
10 & 4  & 8  & 15 & 12 & 1 & 5  & 20 & 11 & 13 \\ \hline
\end{tabular}
\caption{Assignment of 100 products from original categories \textit{A}-\textit{J} to clusters 1-20.}
\label{tabDataAss20}
\end{center}
\end{table}

\begin{table}
\begin{center}
\begin{tabular}{lc}
\hline
Number of classes & 20 \\ 
Purity     & 0.820        \\
Reverse purity      & 0.510        \\
Rand index        & 0.931        \\
Cost function value       & 0.0036        \\ \hline
\end{tabular}
\caption{Evaluation statistics for model with 20 clusters.}
\label{tabDataEval20}
\end{center}
\end{table}

\subsection{Computational Complexity}

\begin{figure}
\begin{center}
\includegraphics[width=11cm]{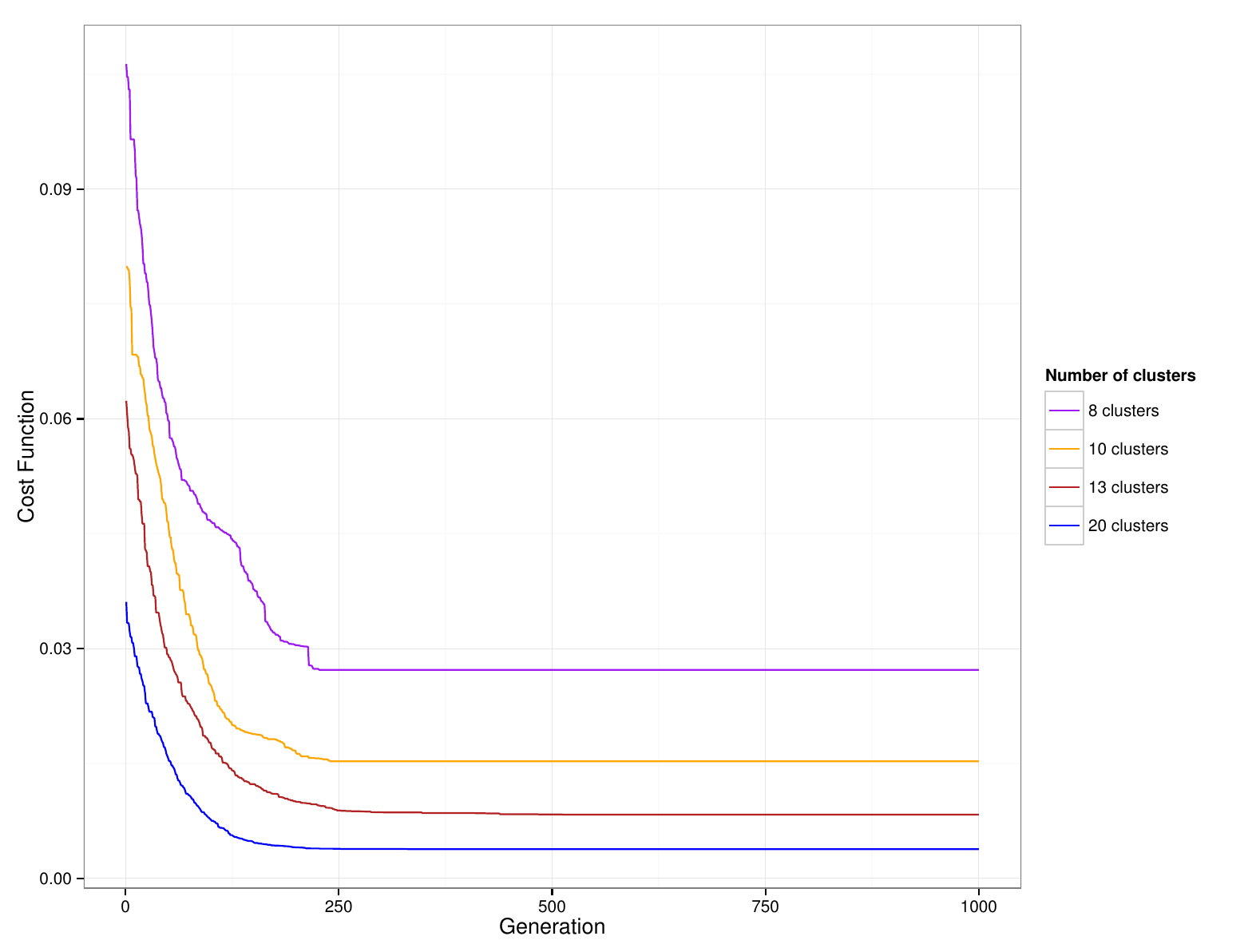} 
\caption{Cost function of best individuals in each generation in models with 8, 10, 13 and 20 clusters.}
\label{figGenerationReal}
\end{center}
\end{figure}

The behaviour of our implementation of the genetic algorithm is regular. Regarding actual time, it took approximately two hours to finish 1000 generations using common PC (i7 CPU with 4 cores). It seems that it is not needed to include such a large number of generations.
As can be seen in Figure \ref{figGenerationReal}, the final assignment is found approximately by the 400th generation of our genetic algorithm using our dataset based on real data. The rest of the computation was not needed. As we use the genetic algorithm to minimise the cost function and we do not know optimal solution beforehand; we cannot prove that the solution is indeed optimal.   To reduce computatinal time, it may be useful to stop the algorithm after a given number of generations without improvement, e. g. stop the computation if the 50 iterations do not improve the cost function and call the current solution as final. In our cases this would greatly reduce the computational time without affecting the final solution.

%Rarely are products from different categories clustered into one cluster. 

\subsection{Summary}

\begin{figure}
\begin{center}
\includegraphics[width=11cm]{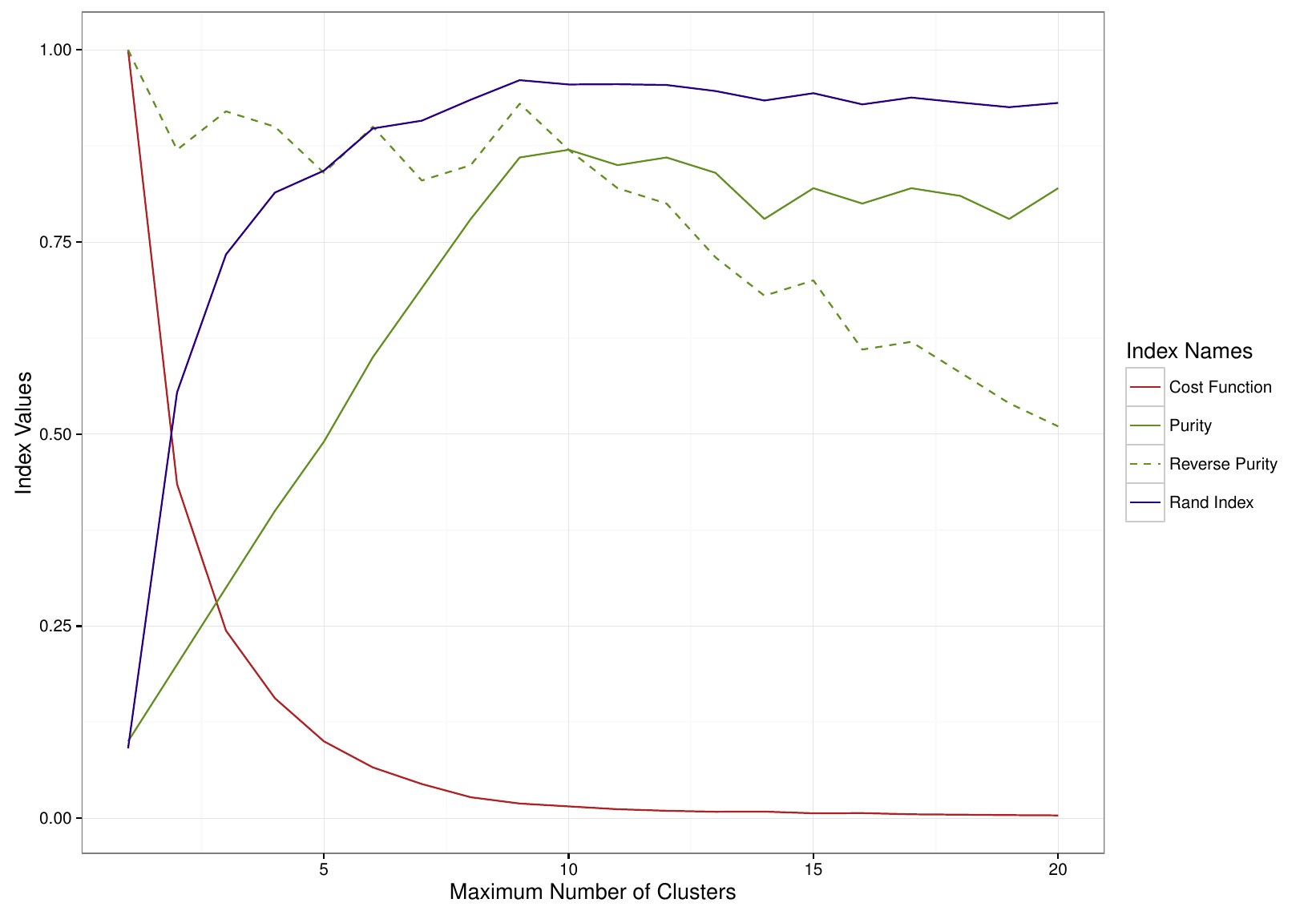} 
\caption{Evaluation statistics for different number of clusters using proposed method.}
\label{figNumberReal}
\end{center}
\end{figure}

The evaluation statistics depend on the number of clusters. Dependency on the number of clusters on real data is similar to the one presented on simulated data in Section \ref{secSimNumber} as can be seen in Fig. \ref{figNumberReal}.

We have found that the product categorization of retail chain is not perfect. The proposed method was able to find clusters which lead to interesting subcategories. That may be used for example in choosing products which are sold in small stores where space on shelves is limited.

Splitting up categories into more exclusive clusters can help with organising the shelves, e. g. not ordering products by the brand but by the other characteristic (which may be found by our method) while the products are in the same category. We remind that in this is dataset we tried to find what were the reasons that made clusters, that may not be necessary needed in the real business with sufficient amount of data. 

In order to maximise utility, the results that are obtained by using our method should be combined with other method, such as categorising of products by function, brand or price appeal. 

The problem we have encountered is that common drugstore's basket contains only a few items while drugstore's assortment is usually much larger than supermarket's. If we have more data the method would give better results. Therefore, we expect our method will work better on supermarket's market basket data with larger amount of different items in the basket and also with thiner range of asortment.
 
%MANAGERIAL INSIGT
%
%Retail companies usually inspect affinity relationship between single products, e. g. sales in same basket normalized by total sales. Clustering of products based solely on market basket data in this area is not so common. It can help mainly in organizing shelf and/or maximizing effect of promotional activities such as newsletter promotions with significant discount. This kind of promotion should attract customers who do not regularly visit the store.
%For example the promotion of two products from the same cluster is not effective as customer usually buys only one of them. This may be helpful mainly in markets with high proportion of sales in promotion, such as Czech Republic drugstore market where over half of sales is in promotion.

\subsection{Comparison with other methods}

%As an applied paper, the natural choice would be k-means, so the authors should convince the reader why their proposal is a better choice for the problem discussed here.

We compared the evaluation statistics of our method with basic method -- in particular $k$-means and Ward's hierarchical clustering. 
To use this various methods we had to transform market basket data to \emph{charasteristics} of given products. We count the percentage of occassions in which products are bought together. Therefore, we get square matrix with dimensions of 100 -- number of products in our sample. Hence, during the data transformation there is a loss of information. 

We used implementations of these methods in R, particulary \emph{kmeans} and \textit{hclust} functions from package \emph{stats} and \emph{som} function from so called package. For parameters of $k$-means method we set 1000 starting locations, maximal number of iterations to 1000 and Hartigan-Wong's algorithm. 
In Ward's hierarchical clustering we set basic Euclidean distance, others parameters remain default. These setups gave the best results. 

\begin{figure}
\begin{center}
\includegraphics[width=11cm]{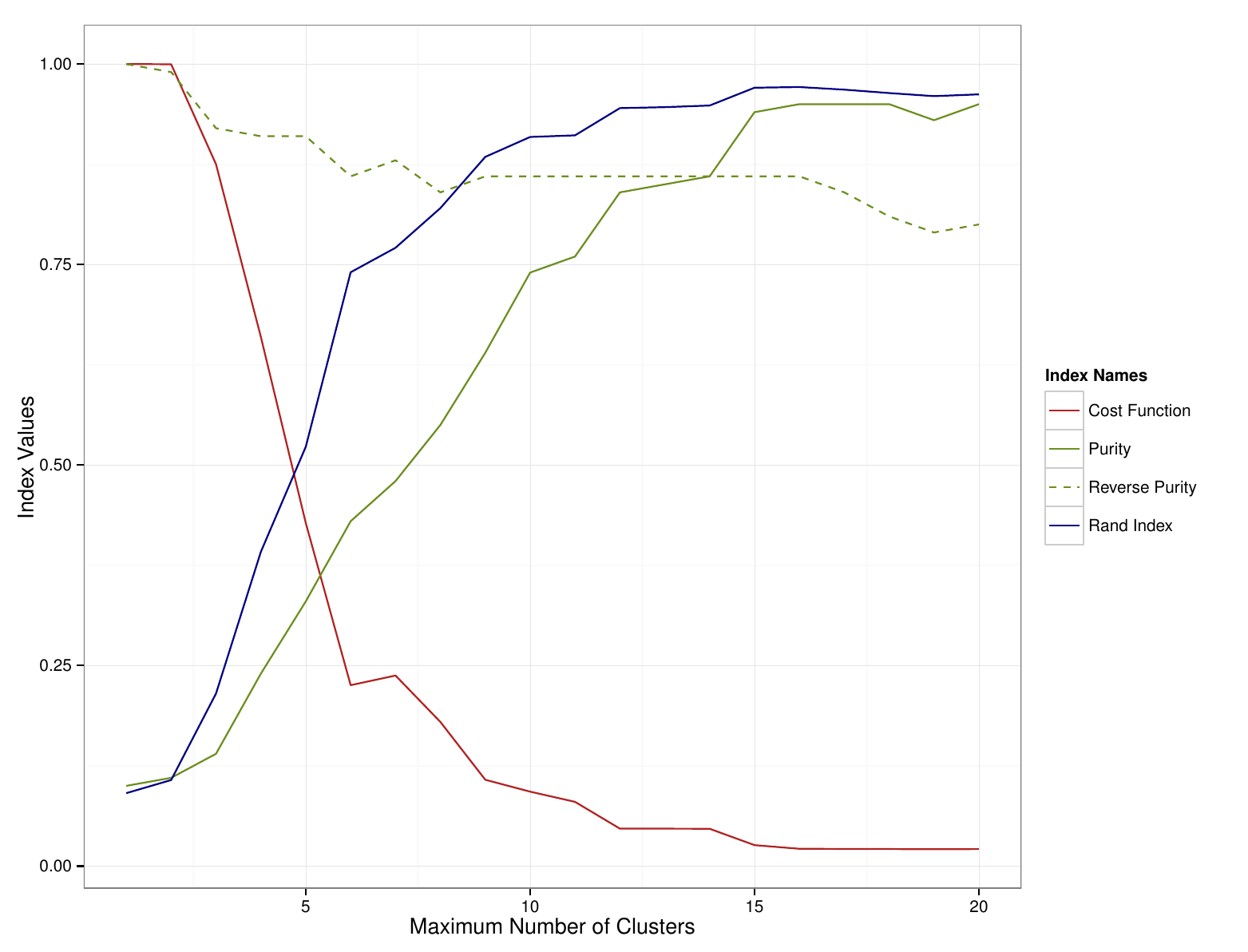} 
\caption{Evaluation statistics for different number of clusters using $k$-means.}
\label{figKmeans}
\end{center}
\end{figure}

The results were interesting. Ward's hierarchical clustering and $k$-means method gave almost identical results in every evaluation statistics. According to evaluation statistics such as purity, reverse purity and Rand index, for lower number of cluster our proposed method gave significantly better results; however, with adding more clusters Ward's hierarchical clustering and $k$-means were more accurate. On the other hand, cost function of Ward's hierarchical clustering and $k$-means is significantly larger for every number of clusters -- as we had to use transformation of data, we could not optimize by our cost function. As a result, the resulting cost function is often more than 10 times larger than in our method.
In Figure \ref{figKmeans} are shown resulting statistics for $k$-means method based on number of clusters. 

It is worth noting that evaluation statistics purity, reverse purity and Rand index are based on the belief that original categories were correctly set (e. g. they fit our assumptions). Only cost function is purely data driven statistic and as we show in Section \ref{secMethClu}, the objective function we proposed should be more appropriate for our goal as we describe in Section \ref{secMethClu}. 

Self-organized map did not work well under two-dimensional setup. We suspect that the the bottleneck of this approach is the dimension-reduction step which is not appropriate method here. Products may not be easily represented in 2D space when our goal is to cluster products which are not commonly bought together.

\section{Conclusion}
\label{secCon}

We introduced a new method for the product categorization based solely on the market basket data. 
The method uses a genetic algorithm for dividing products into a given number of clusters. 

We tested the method using synthetic and real data. The method performs well at synthetic data even if the assumptions are violated to some point.
We verified our method using real market basket data from a drugstore's retail market. We found that the method accurately identified categories which do not significantly violated the assumptions. When the assumption that customers buy at most one product from each category is violated then the products from that category were spread into several clusters instead of assigning to one cluster. It is worth noting that the original categories were subjectively chosen. Our method identified several \textit{hidden} subcategories using only market basket data that may be widely used in marketing and in general in decision-making processes. 

We found out that a common feature of customer's behaviour in the Czech drugstore market is that there are not enough receipts with a larger amount of different products, which lead to a violation of the method's assumptions. If we had more data, we suppose that the method would give even more accurate results. Simulations using synthetic data strongly support this hypothesis.

\section*{Acknowledgements}
\label{sec:acknow}

We would like to thank Miroslav Rada for his kind comments and Alena Holá for proofreading the paper.

\section*{Funding}
\label{sec:fund}

The work of Vladimír Holý and Ondřej Sokol on this paper was supported by IGS F4/63/2016, University of Economics, Prague. The work of Michal Černý was supported by the Czech Science Foundation under Grant P402/12/G097.

%\bibliography{library.bib}
%\bibliographystyle{myjss}

\end{document}